\documentclass[12pt]{iopart}

\providecommand{\openone}{\leavevmode\hbox{\small1\kern-3.8pt\normalsize1}}
\usepackage{iopams}  
\usepackage{graphicx,color}
\begin{document}
\newcommand{\tcr}[1]{\textcolor{red}{#1}}
\title{Semiclassical model for a memory dephasing channel}

\author{D'Arrigo A$^{1}$, Benenti G$^{2,3}$, Falci G$^{1}$}

\address{$^{1}$MATIS CNR-INFM, Catania \&
Dipartimento di Metodologie Fisiche e Chimiche per l'Ingegneria,
Universit\`a di Catania, Viale Andrea Doria 6, 95125 Catania, Italy} 
\address{$^{2}$CNISM, CNR-INFM \& Center for Nonlinear and Complex Systems, 
Universit\`a degli Studi dell'Insubria, Via Valleggio 11, 22100 Como, Italy}
\address{$^{3}$Istituto Nazionale di Fisica Nucleare, Sezione di Milano,
via Celoria 16, 20133 Milano, Italy}

\ead{darrigo@femto.dmfci.unict.it}

\begin{abstract}
We study a dephasing channel with memory, described by a Hamiltonian model in
which the system-environment interaction is described by a stochastic process.
We propose a useful way to describe the channel uses correlations. Moreover, 
we give a general expression for the coherences decay factors as a function of
the number of channel uses and of the stochastic process power spectrum.
We also study the impact of memory on the three qubit code, showing that 
correlations among channel uses affect very little the code performance.
\end{abstract}


\section{Introduction}
State transfer between different units of a quantum computer or 
entanglement distribution between two parties require quantum communication 
channels~\cite{kn:nielsen-chuang,kn:benenti-casati-strini}. They are 
quantum systems transferring quantum information: the proper quantity 
to characterize the channel performance is 
the \textit{quantum capacity}, defined as the maximum number of 
qubits that can be reliably transmitted per channel use~\cite{BennetShor98}.
 
Quantum channels are often thought as memoryless, implying that 
the effect of the channel on each information carrier 
is always described by the same map ${\cal E}$. In other terms 
there is no memory in the interaction between carriers and 
the environmental degrees of freedom physically describing the channel.
In this case the quantum operation for $N$ channel uses is given by 
${\cal E}_N={\cal E}^{\otimes N}$.
 However, in several physically relevant situations this is not a 
 realistic assumption.
 Memory effects appear when the characteristic 
 time scales for the environment dynamics are comparable or longer than the 
time between 
 consecutive channel uses.
 For instance, solid state implementations, which are the most promising 
 for their scalability and integrability, suffer from low frequency 
 noise~\cite{kn:solid_state_environment_noise}. In optical fibers,
 memory effects may appear due to    
 slow birefringence fluctuations~\cite{Banaszek04}. 
 This introduces correlation among uses, then 
 ${\cal E}_N\neq {\cal E}^{\otimes N}$,
 this kind of channels being referred in the literature as {\it memory 
 channels}~\cite{BowenMancini04,MemoryChannel,KretschmannWerner05}.

 A very interesting question, raised for the first time in 
 Ref.~\cite{MacchiavelloPalma02}, is whether memory can {\it enhance}
 the transmission capacity of a quantum channel. 
 Recently we have considered a channel subject to dephasing noise 
 described by a Markov chain, showing that the quantum capacity increases
 with respect to memoryless limit~\cite{DBF_NJP07} 
 (see also~\cite{Hamada,PlenioVirmani07}). Furthermore based on theoretical 
 arguments and numerical simulations, we have conjectured that the enhancement
 of the quantum capacity also takes place for a dephasing quantum environment
 modelled by a bosonic bath~\cite{DBF_NJP07}.
 
 This issue is also relevant for the performance 
 of Quantum Error-Correcting Codes (QECCs). 
 Since quantum capacity is the maximum rate of reliable 
 quantum information transmission, it puts 
 an upper bound to the asymptotic rate achievable by any QECC. 
 On the other hand, realistic 
 QECCs necessarily work on a finite number of channel uses. 
 Moreover, present day experimental implementations~\cite{Cory98,Leung99} 
 are bases on very few channel uses.
 Previous studies have investigated the impact of correlations on the 
 performance of QECCs~\cite{QECCcorrelations,Clemens}. Depending on the 
 chosen model, correlations may have positive or negative impact on QECCs.
 In a previous paper~\cite{DDBF_IJQI08} we have shown, for a Markovian dephasing 
 channel, that also low values of memory, for which the quantum 
 capacity does not change appreciably, can have a detrimental impact on 
 the three-qubit code performance.

 In this paper we describe a dephasing channel by a Hamiltonian where 
 the system environment interaction is modelled by a 
 stochastic process. Then we discuss the three-qubit code 
 error performance in  presence of channel correlations.

\section{Channel Model}
 We suppose that information is carried by qubits that transit across 
 a communication channel, modelled as an environment determining pure 
 dephasing of the qubits. The environment acts as a stochastic drive $\xi(t)$ on the 
 system and the Hamiltonian describing the 
 transmission of $N$ qubits through the channel reads
 \begin{equation}
   {\cal H}(t) = -\,\frac{\lambda}{2}\, \xi(t) \sum_{k=1}^{N}\sigma_z^{(k)}\, f_k(t). 
  \label{Hamiltonian}
 \end{equation}
 The $k-$th qubit is coupled to the environment via 
 its Pauli operator $\sigma_z^{(k)}$, with coupling strength $\lambda$.
 The functions $f_k(t)=u(t-t_k)-u(t-t_k-\tau_p)$, where $u(t)$ is 
	the unit step function~\cite{Abramowitz}, switch the coupling on and off.
 Here $\tau_p$ is  the time each carrier takes to cross the channel; 
 $\tau \equiv t_{k+1}-t_k$ is the time interval that separates
 two consecutive qubits entering the channel.
Only when the $k$-th qubit is inside the channel the function $f_k=1$.
 We assume $\xi(t)$ is a stationary and Gaussian stochastic process~\cite{Papoulis} with 
 zero average value, characterized by its autocorrelation function $C(\tau)$.

 To deal with this problem, we first consider the time evolution of the system 
for a given realization $\xi(t)$
 of the stochastic process, and then we perform an average over all possible realizations. 
 The $N$-qubit time evolution operator for a given realization is
 \begin{eqnarray}
   U_{\xi}(t) \,=\, \bigotimes_{k=1}^{N} \exp(-\rmi \sigma_z^{(k)} \phi_k),
  \label{time_evolution}
 \end{eqnarray}
 where $\phi_k$ is the phase acquired by the $k$-th qubit 
 coherences after the qubit crossed the channel:
 \begin{eqnarray}
       \phi_k\,=\, \frac{\lambda}{2} \int_{t_k}^{t_{k+1}} \hskip-3pt 
	\xi(t^\prime)\, \rmd t^\prime.
 \label{phase_k}
 \end{eqnarray}
 Time evolution is conveniently described in the 
 factorized basis $\{|j\rangle \equiv |j_1,....,j_N\rangle, 
 \,j_1,...,j_N=0,1\}$, where $\{|j_k\rangle\}$ are eigenvectors of
 $\sigma^{(k)}_z$.
 Let $\rho^{\cal Q}=\sum_{j,l} a_{jl} |j\rangle\langle l|$ be the initial state of the $N$-qubit system; the final state $\rho^{\cal Q^\prime}$ after all $N$ qubits crossed the channel 
 is given by
 \begin{eqnarray}
    \rho_{\xi}^{\cal Q^\prime} \,=\, U_{\xi}(t)\, \rho^{\cal Q} U^{\dag}_{\xi}(t)\, \,=\,
   \sum_{j,l} a_{jl}\, \exp\Bigg( 2\rmi\, \sum_{k=1}^{N} s_k\phi_k\Bigg) \,|j\rangle\langle l|,
 \label{rho_out-xi}
 \end{eqnarray}
 where $s_k\equiv l_k-j_k=\frac{1}{2}[(-1)^{j_k}-(-1)^{l_k}]$. 
 By averaging over the stochastic process we finally obtain 
 \begin{eqnarray}
    \rho^{\cal Q^\prime} = \langle \rho_{\xi}^{\cal Q^\prime} \rangle \,=\,
    \sum_{j,l} a_{jl}\, \Bigg\langle \exp\Bigg( 2\rmi\, \sum_{k=1}^{N} s_k\phi_k\Bigg) 
                         \Bigg\rangle \,|j\rangle\langle l|,
 \label{rho_out}
 \end{eqnarray}
which is a quantum operation for the N-qubits system: $\rho^{\cal Q^\prime}={\cal E}_{N}\big(\rho^{\cal Q}\big)$. 
It is possible to show that the quantity $\sum_{k=1}^{N} s_k\phi_k$ is itself
a Gaussian variable\footnote{Given two Gaussian variables $x$ and $y$ with arbitrary 
    variances and some degree of correlation, the two variables $z_{\pm}=x\pm y$ are again 
    Gaussian, as one can find by deriving the $(z_+,z_-)$ mixed density function from the 
    one of $(x,y)$~\cite{Papoulis}. Each phase $\phi_k$ is a time integral of a Gaussian 
    stochastic process, so it can be view as the limit of a sum of Gaussian variables, 
    so in its turn it is Gaussian.},
 therefore 
 \begin{eqnarray}
     \Bigg\langle \exp\Bigg(2\rmi\, \sum_{k=1}^{N} s_k\phi_k \Bigg)\Bigg\rangle \, =\,
     \exp\Bigg(-2 \sum_{k,k^\prime=1}^{N} s_k s_{k^\prime}\langle \phi_k\phi_{k^\prime} \rangle \Bigg).
 \label{average}
 \end{eqnarray}
We call this quantity the $(j,l)$-coherence decay factor,
since it is just the damping experienced 
by the $(j,l)$ system coherence: 
\begin{eqnarray}
D_{jl} \equiv \frac{\langle j | \rho^{\cal Q^\prime} | l \rangle}
                       {\langle j | \rho^{\cal Q}  | l \rangle}
     = \exp\Big(-2 \sum_{k,k^\prime=1}^{N} s_k s_{k^\prime}\langle \phi_k\phi_{k^\prime} \rangle \Big).
 \label{Dmn_def}
 \end{eqnarray}
Next, by using the stationarity of $\xi(t)$, we calculate
 \begin{eqnarray}
  \langle \phi_k\phi_{k^\prime} \rangle = &&
    \Big\langle \frac{\lambda^2}{4} \int_{t_k}^{t_k+\tau_p}\hskip-3pt \rmd t_1 \, \xi(t_1)
                                    \int_{t_{k^\prime}}^{t_{k^\prime}+\tau_p} \hskip-3pt  \rmd t_2 \,\xi(t_2) 
    \Big\rangle \,=\, \nonumber\\ 
             && \frac{\lambda^2}{4} \int_{t_k}^{t_k+\tau_p} \hskip-3pt  \rmd t_1 
                                    \int_{t_{k^\prime}}^{t_{k^\prime}+\tau_p} \hskip-3pt  \rmd t_2\,\,
                                    C(t_1-t_2).
    \label{phase-correlation-1}
 \end{eqnarray}
 Since the autocorrelation function can be expressed in terms of the 
 power spectral density $S(\omega)=\int \rmd \tau \rme^{i \omega \tau} C(\tau)$ of $\xi(t)$ we obtain 
 \begin{eqnarray}
  \langle \phi_k\phi_{k^\prime} \rangle =
    \lambda^2 \int_0^{\infty} \frac{\rmd \omega}{2 \pi} S(\omega) 
                              \frac{1-\cos(\omega \tau_p)}{\omega^2}
                              \cos[\omega(k-k^\prime)\tau],
 \label{phase-correlation-2}
 \end{eqnarray}
and the final form of the coherences decay factor follows: 
 \begin{eqnarray}
    D_{jl} &&= \exp\Bigg(
    -\lambda^2 \int_0^{\infty} \frac{\rmd \omega}{\pi} S(\omega) 
                               \frac{1-\cos(\omega \tau_p)}{\omega^2} \cdot \nonumber \\
           && \hspace{3.5cm}
                             \sum_{k,k^\prime=1}^{N} (l_k-j_k) (l_{k^\prime}-j_{k^\prime}) \cos[\omega(k-k^\prime)\tau] 
                 \Bigg)
 \label{Dmn}
 \end{eqnarray}
This result is identical  to the coherences decay 
due to a dephasing channel modelled by a set 
of quantum harmonic oscillators~\cite{DBF_NJP07}.

 The expression \eref{Dmn} can be put in a nice and useful form. To this end we define
 $\mu_{kk^\prime}$ as the correlation coefficient~\cite{Papoulis} between the phases $\phi_k$ and
 $\phi_{k^\prime}$:
 \begin{eqnarray}
   \mu_{kk^\prime}\,=\,\frac{\langle \phi_k \phi_{k^\prime}\rangle}
                  {\sqrt{\langle\phi^2_k\rangle \langle\phi^2_{k^\prime}\rangle}} 
            \,=\, \frac{\langle \phi_k \phi_{k^\prime}\rangle}{\eta^2},
 \label{phase-correlation-factor}
 \end{eqnarray}
 where we set $\eta^2 = \langle\phi^2_k\rangle$; in fact, thanks to stationarity of
 the process, the quantity $\langle\phi^2_k\rangle$ does not depend on $k$ 
 (see equation  \eref{phase-correlation-2}).
 The coherence decay factor \eref{Dmn_def} can be rewritten as
 \begin{eqnarray}
    D_{jl} = \exp\Big(-2\eta^2 \sum_{k,k^\prime=1}^{N} s_k s_{k^\prime} \mu_{kk^\prime} \Big).
 \label{Dmn_def2}
 \end{eqnarray}
 Now we observe that $g \equiv \exp(-2\eta^2)$ is just the damping experienced by  
 single qubit coherences for one channel use ($N=1$). We finally write
 \begin{eqnarray}
    D_{jl} = g^{\sum_{k,k^\prime=1}^{N} s_k s_{k^\prime} \mu_{kk^\prime} } =
             g^{\big(\sum_{k=1}^{N} s^2_k \,+ \,2\sum_{k^\prime=1, k>k^\prime}^{N} s_k s_{k^\prime} \mu_{k-k^\prime} \big)}.
 \label{Dmn_def3}
 \end{eqnarray}
 where we have defined $\mu_{k-k^\prime} = \mu_{kk^\prime}$.
 In fact the stationarity of $\xi(t)$
 implies that $\mu_{kk^\prime}$ depends only
 on $|k-k^\prime|$. The quantity $\mu_{k-k^\prime}$ is a measure of the 
 degree of the correlation between the channel uses $k$ and $k^\prime$.

\section{Three-Qubit Code performance}
 As a measure of the quantum information transmission reliability
 we use the \textit{entanglement fidelity}~\cite{Schumacher96}.
 To define this quantity we look at the system ${\cal Q}$
 as a part of a larger quantum system ${\cal RQ }$,
 initially in a pure entangled  state $|\psi^{\cal RQ } \rangle$.
 The initial density operator of the system ${\cal Q }$
 is then obtained from that of ${\cal  RQ}$  by a partial trace over 
 the reference system ${\cal R}$:
 $\rho^{\cal Q}={\rm Tr}_{\cal R} [|\psi^{\cal RQ }
 \rangle  \langle \psi^{\cal RQ } | ]$.
 The system $\cal Q$ is sent through the channel, while $\cal R$ remains ideally isolated 
 from any environment, being $\rho^{\cal RQ^\prime}$ the final state of $\cal RQ$ after the 
 transmission. 
 Entanglement fidelity is just the fidelity between the initial and
 the final state of $\cal RQ$:
 \begin{equation}
  F_e=\langle\psi^{\cal RQ}| \rho^{\cal RQ^\prime} |\psi^{\cal RQ}\rangle. 
 \label{Fe}
 \end{equation}
 First we consider a single use of the channel described by  
 Hamiltonian~\eref{Hamiltonian}. We suppose to feed the 
 channel with a quantum source~\cite{Barnum98} described by the density operator 
 $\rho^{\cal Q}=\frac{1}{2} \openone$.
 The entanglement fidelity is~\cite{DBF_EPJ08}:
 \begin{equation}
   F_e\,=\,\langle\psi^{\cal RQ}| 
      {\mathbb I}^{\cal R}\otimes {\cal E}^{\cal Q}
      \big(|\psi^{\cal RQ}\rangle\langle\psi^{\cal RQ}|\big)
           |\psi^{\cal RQ}\rangle \,=\, \frac{1+g}{2},
   \label{oneuse-Fe}
 \end{equation}
 where ${\mathbb I}^{\cal R}$ is the identity operator and 
 ${\cal E}^{\cal Q}={\cal E}_1$.
 This case is relevant in quantum information field as it takes place when two 
 communication parties try to share a Bell state: the party that initially possesses
 the pair sends one half of it through the quantum channel $\cal E$.
 $F_e$ is the fidelity between the actually shared pair and the original one; it
 means that a Bell measurement on $\rho^{\cal RQ^\prime}$ able to distinguish the 
 ideally shared state from the other states of the Bell basis fails with 
 \textit{error probability}
 $P_{\rme} = 1-F_e$. From \eref{oneuse-Fe} it follows that the error probability 
 for a single channel use is $\frac{1-g}{2}$; in what follows we identify this quantity 
 by $\epsilon$.

\begin{figure}[!ht]
 \centering
   \includegraphics[width=101mm,height=37mm]{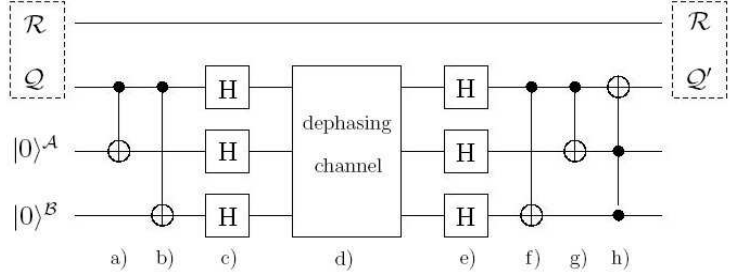}
 \caption{Scheme of a three qubit code~\cite{Cory98}. This quantum error correcting code 
          was initially designed for a bit flip channel, for this reason each channel 
          use (stage d) is embedded between two Hadamard 
          gates~\cite{kn:nielsen-chuang,kn:benenti-casati-strini} (stages c and e).
          The coding is performed by means of CNOT gates (stages a, b, f and g), the 
          decoding also requiring a Toffoli 
          gate~\cite{kn:nielsen-chuang,kn:benenti-casati-strini} (stage h).}
 \label{fig:TQC}
\end{figure}

 Now we suppose to use the Three-Qubit 
 Code (TQC)~\cite{kn:nielsen-chuang,kn:benenti-casati-strini} to send  $\rho^{\cal Q}$.
 The system's state is encoded by using 
 two ancillary systems $\cal A$ and $\cal B$.
 The system and the ancillary qubits are encoded by means of a set of quantum operations 
 that we resume as $\cal C^{\cal QAB}$
 (stages a, b, c in \fref{fig:TQC}) and then transmitted in $N=3$ uses
 of channel~\eref{Hamiltonian}.
 Then the receiver performs the decoding ${\cal D}^{\cal QAB}$ (stages e, f, g, 
 h in \fref{fig:TQC}) on the system $\cal QAB$. After 
 tracing out $\cal AB$, he obtains the final, generally mixed 
 state of system ${\cal RQ}$:
 \begin{equation}
  \rho^{\cal RQ^\prime}_{TQC}=\textrm{tr}_{\cal AB}\big[\mathbb{I}^{\cal R} \otimes 
                 {\cal D}^{\cal QAB} \circ {\cal E}^{\cal QAB} \circ {\cal C}^{\cal QAB} 
                   \big(|{\psi}^{\cal RQAB}\rangle \langle{\psi}^{\cal RQAB}|\big)\big]
 \end{equation}
 where $|{\psi}^{\cal RQAB}\rangle =|{\psi}^{\cal RQ}\rangle\otimes |00^{\cal AB}\rangle $.
 Entanglement fidelity 
 $F_e^{(TQC)}=\langle\psi^{\cal RQ}| \rho_{TQC}^{\cal RQ^\prime} |\psi^{\cal RQ}\rangle$
 just gives the probability that the code is successful.
 The merit of this code is that it drastically reduces - in absence of use correlations, 
 i.e. for ${\cal E}^{\cal QAB}={\cal E}_1^{\otimes 3} $ - 
 the transmission error probability 
 from $\epsilon$ to $P_{\rme}^{(TQC)}=1-F_e^{(TQC)} \simeq 3\epsilon^2$. 

 Now we investigate the effects of channel correlations on the performance of a TQC.
 After some involved calculations it comes out that
\begin{eqnarray}
  F_e^{(TQC,m)} &=& \frac{1}{2} \,+\, \frac{3}{4}g \,-\, \frac{1}{16}g^3 
    \big[
         g^{2\mu_{\cal QA}-2\mu_{\cal QB}-2\mu_{\cal AB}} \,+\, 
         g^{-2\mu_{\cal QA}+2\mu_{\cal QB}-2\mu_{\cal AB}} \,+\,  
         \nonumber\\
      &&\hspace{3.1cm}
         g^{-2\mu_{\cal QA}-2\mu_{\cal QB}+2\mu_{\cal AB}} \,+\, 
         g^{2\mu_{\cal QA}+2\mu_{\cal QB}+2\mu_{\cal AB}}
    \big].
\label{FE-TQC-m-1}
\end{eqnarray}
By observing that $\mu_{\cal QA}=\mu_{\cal AB} = \mu_1$ and 
$\mu_{\cal QB}= \mu_2$ we can rewrite equation \eref{FE-TQC-m-1} as:
\begin{eqnarray}
  F_e^{(TQC,m)} &=& \frac{1}{2} \,+\, \frac{3}{4}g \,-\, \frac{1}{16}g^3 
    \big[2 g^{-2\mu_2} + g^{2\mu_2-4\mu_1} + g^{2\mu_2+4\mu_1} \big].
 \label{FE-TQC-m-2}
\end{eqnarray}
 An interesting result can be obtained by considering the
 case of a small error probability $\epsilon \ll 1$. 
 In this regime we can take the series expansion of \eref{FE-TQC-m-2} 
 near $\epsilon = 0$:
 \begin{eqnarray}
  F_e^{(TQC,m)} \simeq 1\,-\,(3+4\mu_1^2+2\mu_2^2)\epsilon^2.
   \label{FE-TQC-approx}
 \end{eqnarray}
 This expression tells us that even though memory lowers the fidelity, 
 this worsening is always slight and absolutely negligible when 
$\mu_1,\mu_2 \ll 1$. 
 Moreover, it highlights that channel correlations - inside the Hamiltonian model   
 \eref{Hamiltonian} - permit the TQC to maintain its error probability
 $P_{\rme}^{(TQC,m)}=1- F_e^{(TQC,m)}$  of the order of $\epsilon^2$. 
 However, one has to take care that in the case perfect memory the code error
 triplicates from $3\epsilon^2$ to $9\epsilon^2$. 
 These results are very similar to the ones by Clemens {\it et al.}~\cite{Clemens}. However, we discuss time rather than space correlations and average with respect to stochastic processes.
  
\begin{figure}[!tbp]
  \centering
  \includegraphics[width=75mm,height=58mm]{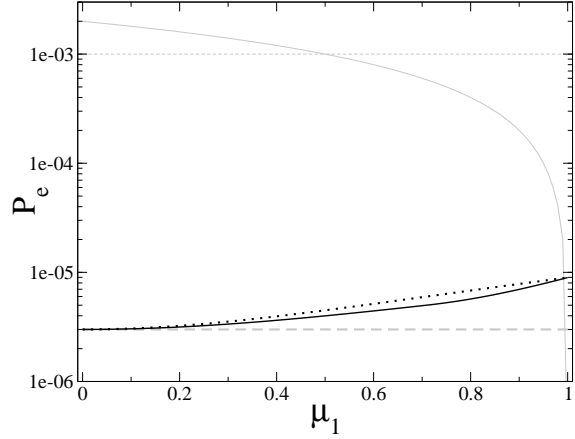}
  \caption{Plot of code error probability as a function of $\mu_1$. The dotted and the dashed 
   grey lines represent respectively the single channel use error probability ($\epsilon=10^{-3}$)
   and TQC error probability for the memoryless channel ($P_\rme^{(TQC)}$). 
   The error probabilities for the three qubit code in presence of correlations 
   ($P_\rme^{TQC, m}$) are represented by black curves: solid curve refers to
   $\mu_2=\max(0,\,\tilde{\mu}_2)$ and the dotted one to $\mu_2=\mu_1$.
   There is also displayed (solid gray curve) the error probability for a simple
   two-qubit code encoding a qubit into the subspace spanned by $\{|01\rangle,|10\rangle\}$.}
 \label{fig:Pe_mu}
\end{figure}

 Rather than choosing a particular autocorrelation function $C(\tau)$ for $\xi(t)$ 
 and then trying to carry out a specific relation between it and the TQC error probability,
 we make some general considerations about the impact of correlations on the code error probability. 
 For the sake of simplicity we assume $\mu_1$ and $\mu_2$ having positive values 
 (we do not consider anti-correlation cases).
 While the range of $\mu_1$ is $[0,1]$, we can argue that $\mu_2 \leq \mu_1$
 since one expects that the phase correlation does not increase when increasing
the channel uses distance;
 furthermore it can be proved that it must be $\mu_2\geq \tilde{\mu}_2\equiv 2\mu_1^2-1$\footnote{One can 
 see it by considering the average of $[\phi_2+a(\phi_1+\phi_3)]^2$ where $a$ is a 
 real variable: it is a quadratic form in $a$ and by imposing that it must always be 
 positive we obtain the desired condition on $\mu_2$~\cite{Papoulis}.}.
 Studying the first derivatives of $P_\rme^{(TQC,m)}$ as a function of $\mu_1$ and $\mu_2$
 it turns out that $P_\rme^{(TQC,m)}$ is monotonical with respect to $\mu_1$ 
 (error grows with $\mu_1$), but not with respect to $\mu_2$. 
 It indeed can displays a minimum at 
 $\mu_{2_{opt}} \equiv -0.25\log_g[(g^{4\mu_1}+g^{-4\mu_1})/2]$, but its presence is 
 substantially irrelevant, and one can reasonably say that the code error probability 
 is also increasing with respect to $\mu_2$. 
 Thus to characterize the $P_\rme^{(TQC,m)}$ behaviour, we plot it as a function 
 $\mu_1$, using $\mu_2\in \{\max(0,\tilde{\mu}_2),\,\mu_1\}$  as parameter.
 As it is showed in \fref{fig:Pe_mu}, in which we set $\epsilon=10^{-3}$, the TQC error 
 probability weakly depends on $\mu_1$, and the $\mu_2$ allowable values affect very 
 little it. In the same figure we also plot the error probability
 for a two-qubit code~\cite{DBF_EPJ08} encoding a qubit into the subspace spanned by 
 $\{|01\rangle,|10\rangle\}$: 
 the performance of this last code is always worse than the TQC ones, unless in the case
 $\mu_1 \to 1$, for which the coding subspace becomes decoherence-free.

\begin{figure}[!ht]
 \centering
 \includegraphics[width=75mm,height=58mm]{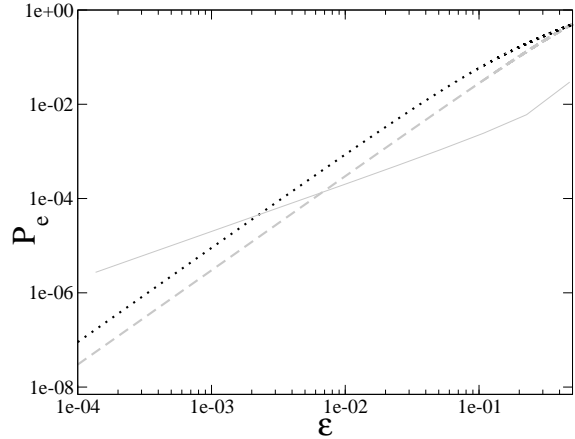}
 \caption{Plot of code error probability as a function of the single channel use
  error probability $\epsilon$. The dashed 
  grey line represents the TQC error probability in the memoryless case ($P_\rme^{(TQC)}$). 
  For the error probabilities of the three qubit code in presence of correlations 
  ($P_\rme^{(TQC, m)}$) we plot the worst case $(\mu_2=\mu_1=1)$ by a dotted black curve.
  There is also displayed (solid gray curve) the error probability for a simple 
  two-qubit code encoding a qubit into the subspace spanned by 
  $\{|01\rangle,|10\rangle\}$ for $\mu_1=0.99$ 
  (triangles down).}
 \label{fig:Pe_e}
\end{figure}

 The TQC exhibits the same kind of behaviour showed in \fref{fig:Pe_mu}
 as $\epsilon$ changes. In \fref{fig:Pe_e} we plot $P_\rme^{(TQC, m)}$ as a function
 of $\epsilon$ for $\mu_2=\mu_1=1$, the case in which the code shows the worst 
 performance; we do not plot the cases of low correlations $(\mu_1\leq 0.1)$
 since the correspondent curves are practically indistinguishable from the memoryless ones.
 We also compare $P_\rme^{(TQC, m)}$ with the error 
 probability of the two-qubit code~\cite{DBF_EPJ08}: to produce good results
 this last one requires very high degrees of correlation between successive channel uses.

 In conclusion we find that a Hamiltonian formulation of a memory dephasing channel
 shows that the three qubit code is robust against channel correlations. 
 Significantly different results
 emerge if we describe channel correlations inside a Markovian model~\cite{DDBF_IJQI08}:
 In this latter case memory restores the $\epsilon-$dependence in the code error probability, 
 thus drastically reducing the code performance.

\ack{We acknowledge Andrea Mastellone and Elisabetta Paladino for 
 invaluable help.
 A.D. and G.F. acknowledge support from the EU-EuroSQIP (IST-3-015708-IP) and
 MIUR-PRIN2005 (2005022977).}


\section*{References}
\begin{thebibliography}{10}

\bibitem{kn:nielsen-chuang} Nielsen M A and Chuang I L,
	 \textit{Quantum Computation and Quantum Information}
 	 (Cambridge Unuversity Press, Cambridge 2000)
\bibitem{kn:benenti-casati-strini} 
         Benenti G, Casati and G Strini G 
         {\it Principles of Quantum Computation and Information},
         Vol. I: Basic concepts (Singapore: World Scientific, 2004);
         Vol. II: Basic tools and special topics
         (Singapore: World Scientific, 2007).
\bibitem{BennetShor98} Bennet C H and Shor P 1998 {\it IEEE Trans. Inf. Theory}
         \textbf{44} 2724.
\bibitem{kn:solid_state_environment_noise}
         Makhlin Y {\it et al.} 2001 {\it Phys. Rev. Mod.} \textbf{73} 357;
         Paladino E {\it et al.} 2002 {\it Phys. Rev. Lett.} {\bf 89} (2005) 228304;
         Falci G {\it et al.} 2005 {\it Phys. Rev. Lett.} {\bf 94} 167002;
         Ithier G {\it et al.} 2005 {\it Phys. Rev. B} {\bf 72} 134519.
\bibitem{Banaszek04} Banaszek K {\it et al.}. 2004 
  	 {\it Phys. Rev. Lett} {\bf 92} 257901.
\bibitem{BowenMancini04} Bowen G and Mancini S 2004
         {\it Phys. Rev. A} {\bf 69} 012306; 
\bibitem{MemoryChannel}
         Giovannetti V 2005 {\it J. Phys. A: Math. Gen.} {\bf 38} 10989;
         Datta N and Dorlas T C 2007 J. Phys. A: Math. Theor. 40 8147-8164.         
\bibitem{KretschmannWerner05} Kretschmann D and Werner R F 2005
         {\it Phys. Rev. A} {\bf 72} 062323.           
\bibitem{MacchiavelloPalma02} Macchiavello C and Palma G M 2002
	 {\it Phys. Rev. A} {\bf 65} 050301.
\bibitem{DBF_NJP07} D´Arrigo A, Benenti G, and Falci G 2007 \NJP \textbf{9} 310 
\bibitem{Hamada} Hamada H 2002 
         {\it J. Math. Phys.} {\bf 43} 4382.  
\bibitem{PlenioVirmani07} Plenio M B Virmani S 2007 {\it Phys. Rev. Lett.} 
         {\bf 99} 120504.
\bibitem{Cory98} Cory D G {\it et al.} 1998
	 {\it Phys. Rev. Lett.} {\bf 81} 2152. 
\bibitem{Leung99} Leung D {\it et al.} 1999
	 {\it Phys. Rev. A} {\bf 60} 1924. 
\bibitem{QECCcorrelations} 
         Klesse R and Frank S 1996
         {\it Phys. Rev. Lett.} {\bf 95} 230503; 
         Duan L-M and Guo G-C 1999
         {\it Phys. Rev. A} {\bf 59} 4058; 
         Novais E {\it et al.} 2006 
         {\it Phys. Rev. Lett.} {\bf 97} 040501; 
         Shabani A 2008 {\it Phys. Rev. A} \textbf{77} 022323.
\bibitem{Clemens} Clemens J P {\it et al.} 2004 
         {\it Phys. Rev. A} {\bf 69} 062313;
\bibitem{DDBF_IJQI08} D´Arrigo A, De Leo E, Benenti G, and Falci G 2008 
        {\it IJQI} \textbf{160} 83
\bibitem{Abramowitz} Abramovitz M and Stegun I 1968 {\it Handbook of Mathematical functions} 
         (New York: Dover Publications, Inc.)
\bibitem{Papoulis} Papoulis A 1965 \textit{Probability, Random Variables and
         Stochastic Processes}, McGraw Hill, New York
\bibitem{Schumacher96} Schumacher B 1996 {\it Phys. Rev. A} \textbf{54} 2614 
\bibitem{Barnum98} Barnum H, Nielsen M A and Schumacher B 1998 {\it Phys. Rev. A}
         \textbf{57} 4153. 
\bibitem{DBF_EPJ08} D´Arrigo A, Benenti G, and Falci G 2008 {\it EPJ - ST} \textbf{160} 83

\end{thebibliography}
\end{document}